\documentclass[10pt]{article}
\usepackage{graphicx}

\def\Title#1{\begin{center} {\Large #1 } \end{center}}
\def\Author#1{\begin{center}{ \sc #1} \end{center}}
\def\Address#1{\begin{center}{ \it #1} \end{center}}

\newcommand\pubblock{\rightline{\begin{tabular}{l} Proceedings of the Second Annual LHCP\\
         \pubdate  \end{tabular}}}

\newenvironment{Abstract}{\begin{quotation} \begin{center} 
             \large ABSTRACT \end{center}\bigskip 
      \begin{center}\begin{large}}{\end{large}\end{center} \end{quotation}}

\newenvironment{Presented}{\begin{quotation} \begin{center} 
             PRESENTED AT\end{center}\bigskip 
      \begin{center}\begin{large}}{\end{large}\end{center} \end{quotation}}

\def\beq{\begin{equation}}
\def\eeq#1{\label{#1}\end{equation}}
\def\eeqn{\end{equation}}

\def\beqa{\begin{eqnarray}}
\def\eeqa#1{\label{#1}\end{eqnarray}}
\def\eeqan{\end{eqnarray}}

\let\bar=\overbar

\def\Dslash{\not{\hbox{\kern-4pt $D$}}}
\def\dslash{\not{\hbox{\kern-2pt $\del$}}}

\def\msb{{\bar{\ssstyle M \kern -1pt S}}}

\usepackage{color}
\usepackage{lineno}
\usepackage{hyperref}

\newcommand\pubdate{\today}

\def\affiliation{
On behalf of the ALICE Experiment, \\
Institute for Subatomic Physics \\
Utrecht University,  Utrecht, 3584 CC, The Netherlands }

\begin{document}

\large
\begin{titlepage}
\pubblock

\vfill
\Title{Measurements of heavy-flavour production and azimuthal anisotropy in Pb--Pb collisions with the ALICE detector. }
\vfill

\Author{ Andrea Dubla  }
\Address{\affiliation}
\vfill

\begin{Abstract}

Hadrons containing heavy quarks, i.e. charm or beauty, are unique probes of the properties of the hot and dense QCD medium produced in heavy-ion collisions. Due to their large masses, heavy quarks are produced at the initial stage of the collision, almost exclusively via hard partonic scattering processes. Therefore, they are expected to experience the full collision history propagating through the QCD medium losing energy via elastic and inelastic collisions with the medium constituents.
The ALICE collaboration has measured the production of open heavy-flavour hadrons via their hadronic and semi-electronic decays at mid-rapidity and in the semi-muonic decay channel at \mbox{forward rapidity} in pp, p--Pb and Pb--Pb collisions. 
In this talk the latest results on the open \mbox{heavy-flavour nuclear modification factor, $R_\mathrm{AA}$, and elliptic flow, $v_{2}$, are presented}.

\end{Abstract}
\vfill

\begin{Presented}
The Second Annual Conference\\
 on Large Hadron Collider Physics \\
Columbia University, New York, U.S.A \\ 
June 2-7, 2014
\end{Presented}
\vfill
\end{titlepage}
\def\thefootnote{\fnsymbol{footnote}}
\setcounter{footnote}{0}
%

\normalsize 


\section{Introduction}
The main goal of the ALICE \cite{ALICE2} experiment is to study strongly interacting matter in the conditions of high density and temperature reached in colliding ultra-relativistic lead ions at the LHC. In such collisions a deconfined state of quarks and gluons, the Quark-Gluon Plasma (QGP), is expected to be formed.
Heavy quarks, abundantly produced at LHC energies, are regarded as effective probes of the system formed in nucleus-nucleus collisions. In particular, they should be sensitive to the energy density, through the mechanism of in-medium energy loss.

The nuclear modification factor $R_\mathrm{AA}(p_\mathrm{T}) = \frac{1}{<T_\mathrm{AA}>} \frac{dN_\mathrm{AA}/d\it{p}_{T}}{d\sigma_\mathrm{pp}/{dp_\mathrm{T}}}$ of the $p_\mathrm{T}$ distribution of heavy-flavour hadrons (and their decay leptons) is well established as a sensitive observable for the study of the interaction of hard partons with the medium.
This factor is defined as the ratio of the $p_\mathrm{T}$-differential yield measured in nucleus-nucleus (AA) collisions in a given centrality class to the yield calculated from the proton-proton cross section scaled by the nuclear overlap function $<T_\mathrm{AA}>$, which is proportional to $<N_\mathrm{coll}>$, for that centrality class. The latter is obtained from Glauber model calculations of the collision geometry \cite{Glauber}.
Due to the QCD nature of parton energy loss, quarks are predicted to lose less energy than gluons (that have a larger colour coupling factor) and, in addition, the dead-cone effect \cite{DeadCone} is expected to reduce the energy loss of massive quarks with respect to light quarks.
Therefore, a hierarchy in the $R_\mathrm{AA}$ is expected to be observed when comparing the mostly gluon-originated light-flavour hadrons (e.g. pions) to D and to B mesons \cite{BMe}: \mbox{$R^{\pi}_\mathrm{AA}$ $<$ $R^{D}_\mathrm{AA}$ $<$ $R^{B}_\mathrm{AA}$}. The measurement and comparison of these different medium probes provides a unique test of the color-charge and mass dependence of parton energy loss. A full understanding of these phenomena requires also to quantify the initial-state effects inherent to nuclear collisions, like $k_\mathrm{T}$ broadening and nuclear PDF shadowing \cite{shadowing}. These effects are not related to the QGP phase but to cold nuclear matter and they are studied by measuring the modification of heavy-flavour hadron yields in proton-nucleus with respect to proton-proton collisions ($R_\mathrm{pA}(p_\mathrm{T})$).
Further insight into the medium properties is provided by the measurement of the anisotropy in the azimuthal distribution of particle momenta. 
In heavy-ion collisions the anisotropy in the spatial distribution of the nucleons participating in the collision is converted into a momentum anisotropy, if sufficient rescatterings with the medium constituents occur. Hence the azimuthal distribution of the final-state particles reflects the initial anisotropy and the medium characteristics.
The azimuthal anisotropy of produced particles is characterized by the second Fourier coefficient \mbox{$v_{2} = < \cos[2(\varphi - \psi_{2})] >$}, where $\varphi$ is the azimuthal angle of the particle momentum, and $\psi_{2}$ is the azimuthal angle of the initial-state symmetry plane for the second harmonic \cite{Flow}. 
At low $p_\mathrm{T}$, the $v_{2}$ of heavy-flavour hadrons is sensitive to the degree of thermalization of charm and beauty quarks in the deconfined medium. At higher $p_\mathrm{T}$, the measurement of $v_{2}$ carries information on the path-length dependence of in-medium parton energy loss.
The measurement of heavy-flavour $v_{2}$ offers a unique opportunity to test whether also quarks with large mass participate in the collective expansion dynamics and possibly thermalize in the QGP.

Charm and beauty production was measured with ALICE in Pb--Pb collisions at $\sqrt{s_\mathrm{NN}}$  = 2.76 TeV using electrons and muons from semi-leptonic decays of heavy-flavour hadrons and fully reconstructed D-meson hadronic decays. 
D mesons were reconstructed at mid-rapidity ($|y|$ $<$ 0.5) via their hadronic decay channels: $D^{0} \rightarrow K^{-}\pi^{+}$ (with branching ratio, BR = 3.88$\pm$0.05\%), $D^{+} \rightarrow K^{-}\pi^{+}\pi^{+}$ (BR = 9.13 $\pm$ 0.19\%), $D^{*+} \rightarrow D^{0}\pi^{+}$ (BR = 67.7 $\pm$ 0.5\%) and $D^{+}_{s} \rightarrow \phi \pi^{+} \rightarrow K^{-}K^{+}\pi^{+}$ (BR = 2.28 $\pm$ 0.12\%) and their charge conjugates. 
Due to their large lifetime (c$\tau$=123 $\mu$m, 312 $\mu$m for $D^{0}$ and $D^{\pm}$ respectively), the D mesons do not decay at the primary vertex. The high-precision tracking provided by the Inner Tracking System (ITS) and the Time Projection Chamber (TPC) are used to reconstruct the displaced secondary vertices.
To further reduce the combinatorial background, the decay $\pi^{\pm}$ and $K^{\pm}$ are identified using the measurement of the specific energy loss d{\it E}/d{\it x} in the gas of the TPC and the time of flight of the particle from the interaction point to the Time-Of-Flight (TOF) detector.
The electron identification in the mid-rapidity region \mbox{ ($|y|$ $<$ 0.8)} was based on the d{\it E}/d{\it x} in the TPC. In the low $p_\mathrm{T}$ intervals ($p_\mathrm{T}$ $<$ 3 GeV/{\it c}), where the $K^{\pm}$, proton and deuteron Bethe-Bloch curves cross that of the electron, the measured time-of-flight in TOF and the energy loss in the ITS were employed in addition. At higher $p_\mathrm{T}$, the ratio of the energy deposited in the ElectroMagnetic Calorimeter (EMCal) and the momentum measured with the TPC and ITS, which is close to unity for $\mathrm{e}^{\pm}$, was used to further reject hadrons. 
Muon tracks were reconstructed in minimum bias trigger events in the Forward Muon Spectrometer \mbox{(-4 $<$ y $<$ -2.5)}.


The reference pp cross section of heavy-flavour decay electrons and D mesons at $\sqrt{s_\mathrm{NN}}$  = 2.76 TeV was obtained by a pQCD-based energy scaling of the $p_\mathrm{T}$-differential cross sections measured at $\sqrt{\it{s}}$  = 7 TeV. The heavy-flavour decay muon $R_\mathrm{AA}$ was calculated using the pp cross section measured at $\sqrt{\it{s}}$  = 2.76 TeV \cite{Muonpp}.  
The data analyzed were recorded with the ALICE detector during the 2010 and 2011 LHC runs with Pb--Pb collisions at $\sqrt{s_\mathrm{NN}}$ = 2.76 TeV. The VZERO detector provides the MB trigger, centrality determination, which is used online to enhance the sample of central and semi-central events, and event plane determination. In addition an event sample triggered with the ElectroMagnetic Calorimeter (EMCal) was used to enrich the recorded sample of heavy-flavour decay electrons.

\section{Nuclear modification factor and azimuthal anisotropy}

\begin{figure}[htb]
 \begin{minipage}[b]{0.4\linewidth}
\centering
\includegraphics[height=2.6in]{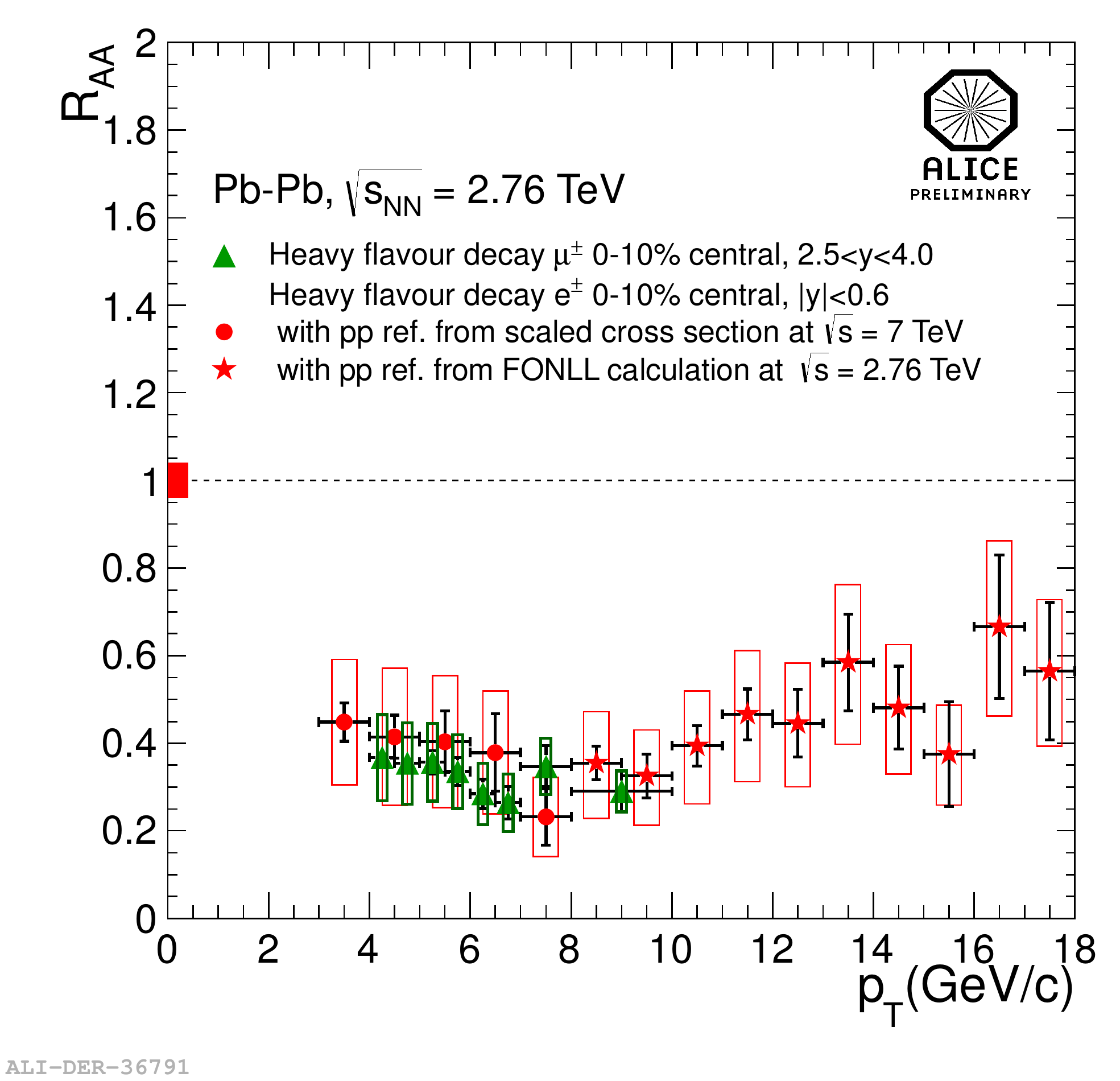}
\end{minipage}
\hspace{0.1cm}
\begin{minipage}[b]{0.5\linewidth}
\centering
\includegraphics[height=2.6in]{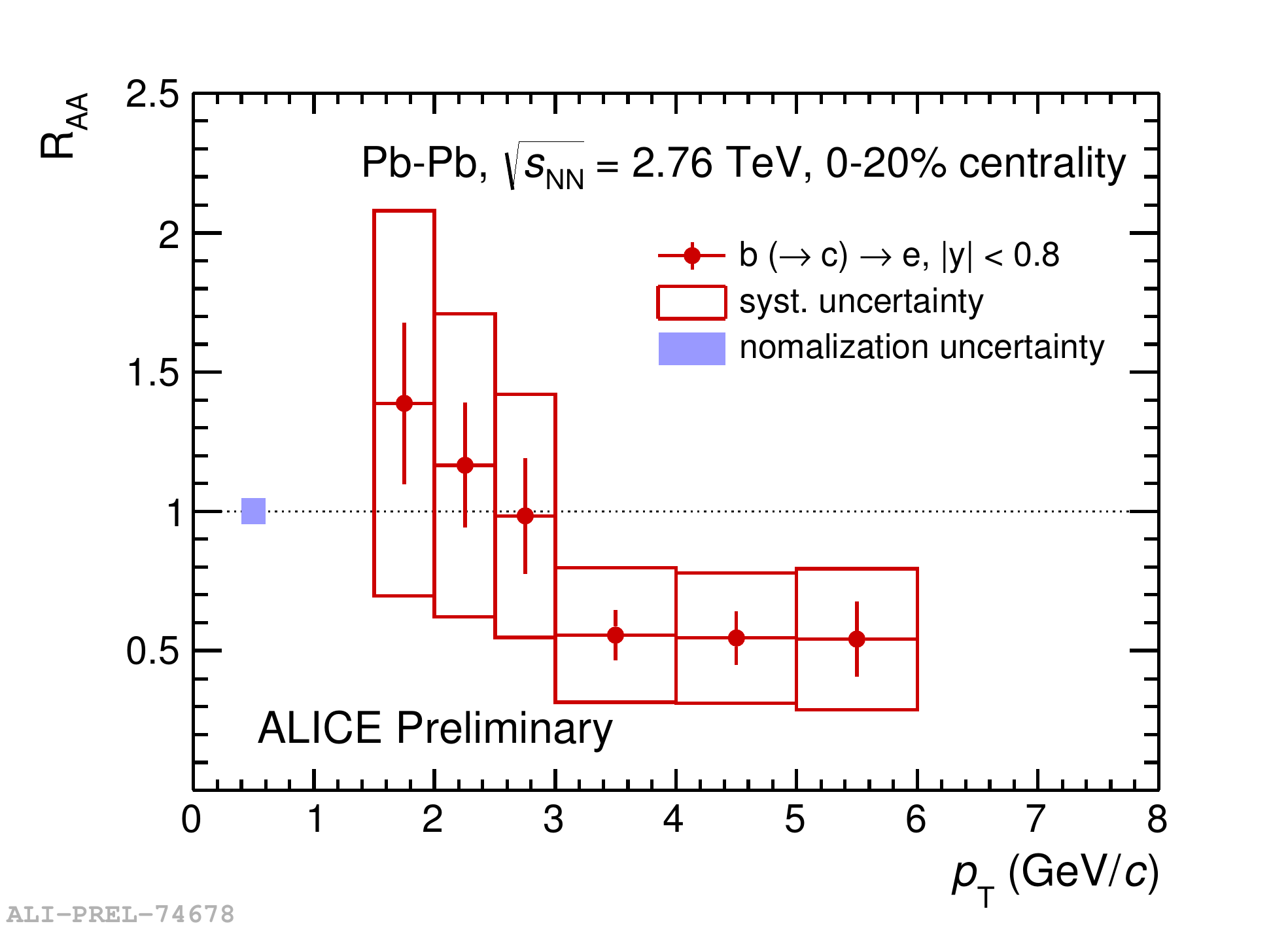}
\end{minipage}
\caption{ Left:  $R_\mathrm{AA}$ of heavy-flavour decay electrons at mid-rapidity and heavy-flavour decay muons at forward rapidity in central (0-10\%) Pb--Pb collisions, as a function of $p_\mathrm{T}$. Right: beauty-decay electron  $R_\mathrm{AA}$ as a function of $p_\mathrm{T}$ measured in central (0-20\%) Pb--Pb collisions.
}
\label{fig:figure1}
\end{figure}

The left panel of Figure \ref{fig:figure1} shows the $R_\mathrm{AA}$ of heavy-flavour decay electrons at mid-rapidity and of heavy-flavour decay muons at forward rapidity, measured in the 10\% most central Pb--Pb collisions. A clear suppression is observed for both electrons and muons in the measured $p_\mathrm{T}$ range and it is compatible within uncertainties at central and forward rapidity.
The contribution to the heavy-flavour decay electron yield due to beauty-hadron decays was extracted by means of a fit to the electron impact parameter distribution. The electron sources were included in the fit through templates obtained from simulations. The measurement shown in the right panel of Figure \ref{fig:figure1} indicates that the beauty-decay electron $R_\mathrm{AA}$ is smaller than unity for $p_\mathrm{T}$ $>$ 3 GeV/{\it c}.
ALICE also measured the nuclear modification factor ($R_\mathrm{pPb}$) of heavy-flavour decay leptons in minimum-bias p--Pb collisions at $\sqrt{s_\mathrm{NN}}$ = 5.02 TeV. The results are compatible with unity \cite{MuonPpb} indicating that the suppression observed in central Pb--Pb collisions is due to final state effects, i.e. the interaction of heavy quarks with the hot and dense medium.

\begin{figure}[htb]
 \begin{minipage}[b]{0.5\linewidth}
\centering
\includegraphics[height=2.75in]{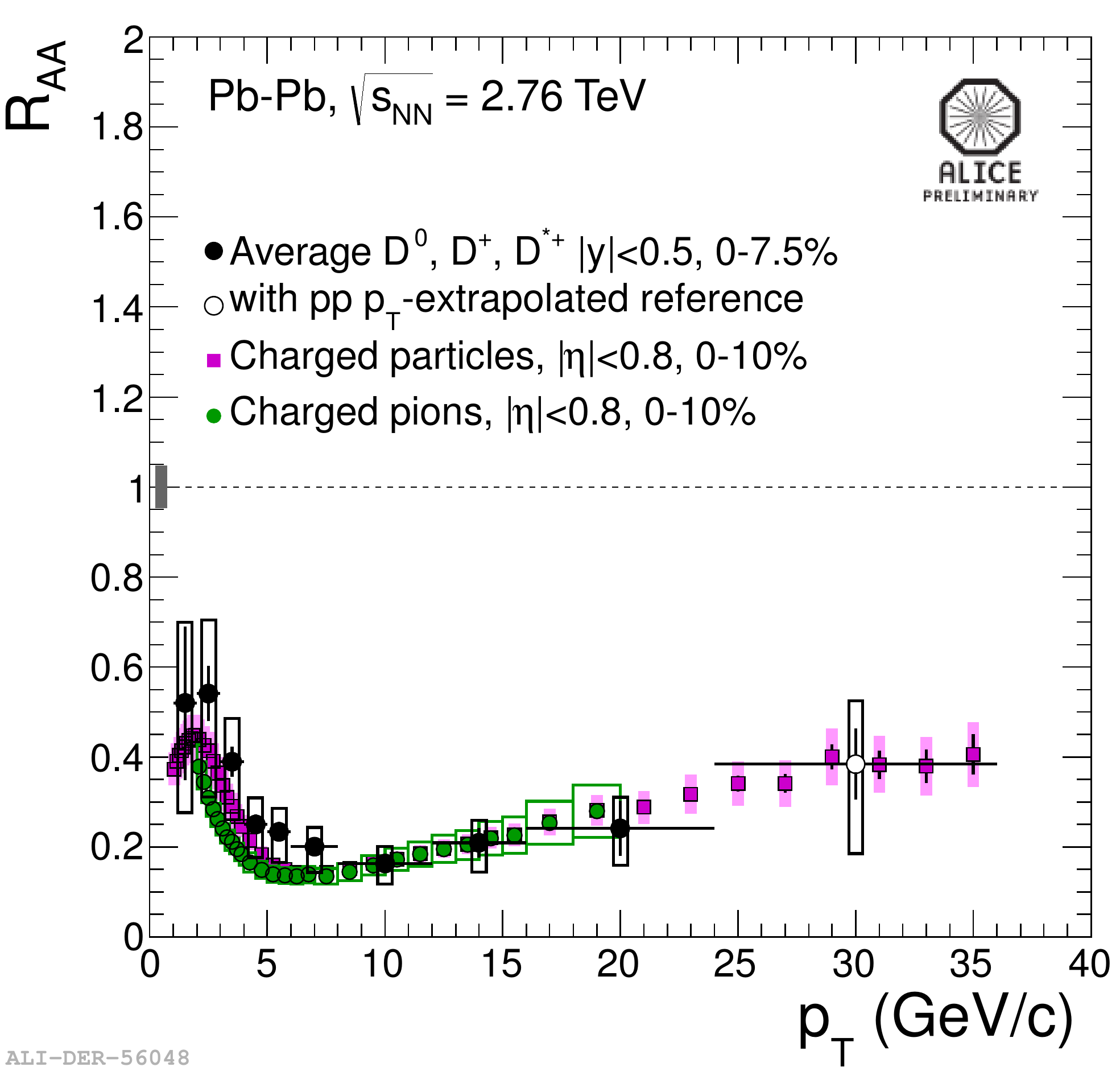}
\end{minipage}
\hspace{0.2cm}
 \begin{minipage}[b]{0.5\linewidth}
\centering
\includegraphics[height=2.75in]{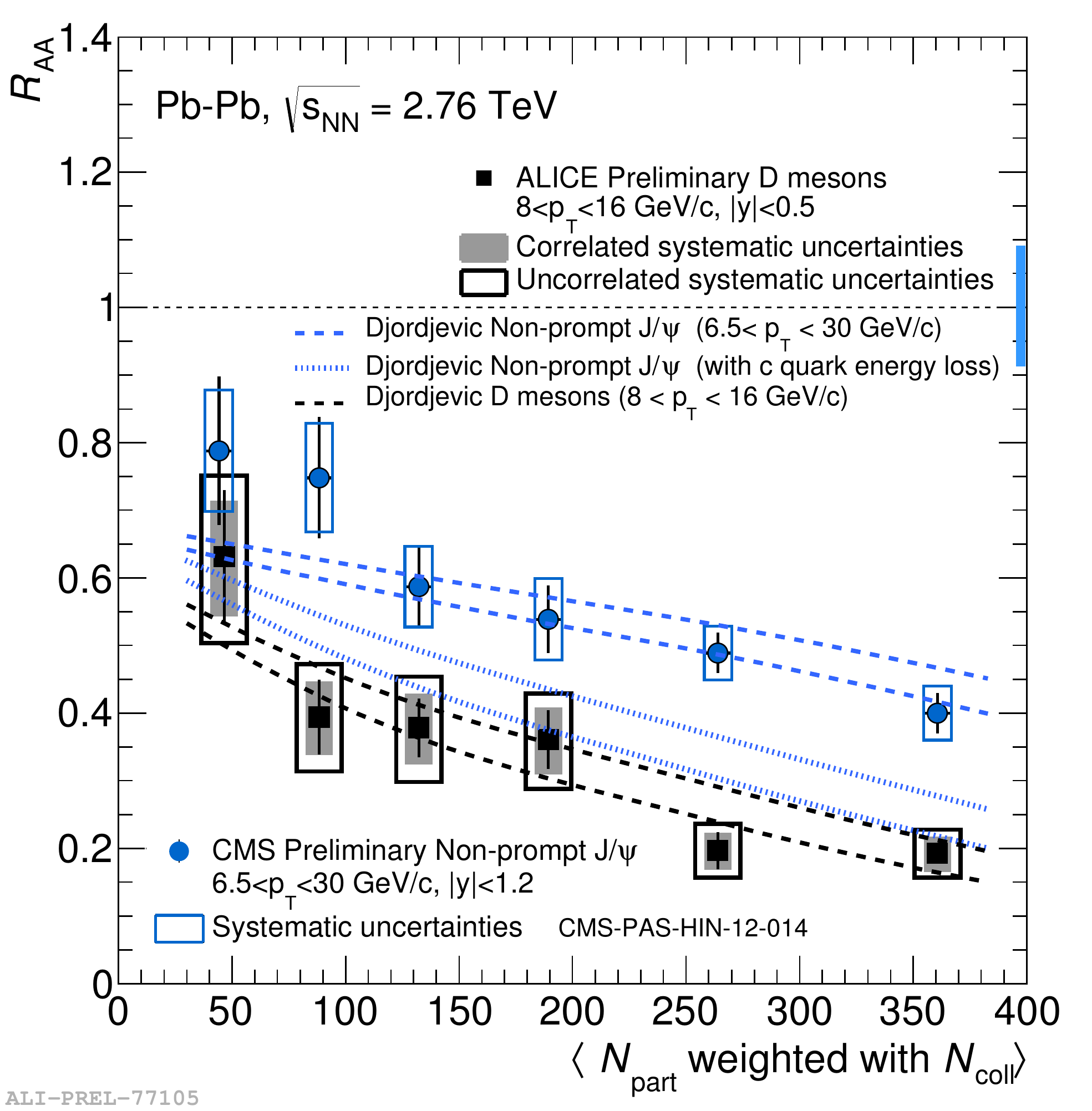}
\end{minipage}
\label{fig:figure2}
\caption{D-meson $R_\mathrm{AA}$ as a function of $p_\mathrm{T}$ in central (0-7.5\%) collisions compared to charged-particle and pion $R_\mathrm{AA}$ measured by ALICE (left panel) and as a function of centrality in the transverse momentum region 8 $<$ $p_\mathrm{T}$ $<$ 16 GeV/{\it c} compared to the J/$\Psi$ from B decays measured by CMS (right panel). The theoretical calculation of D mesons and non-prompt J/$\Psi$ $R_\mathrm{AA}$ by Djordjevic et al. \cite{Djordjevic} including radiative and collisional energy loss are shown as well.}
\end{figure}

The left panel of Figure \ref{fig:figure2} shows the nuclear modification factor of prompt D mesons (average of $D^{0}$, $D^{+}$ and $D^{*+}$) as a function of $p_\mathrm{T}$ in central (0-7.5\%) Pb--Pb collisions. A strong suppression is observed in the measured $p_\mathrm{T}$ range, reaching a factor of about 5 for $p_\mathrm{T}$ $\sim$ 8-10 GeV/{\it c}. The D-meson $R^{D}_\mathrm{AA}$ is compatible with that of charged pions and charged particles within uncertainties \cite{RAACH}, although there seems to be a tendency for $R^{\pi}_\mathrm{AA}$ $<$ $R^{D}_\mathrm{AA}$ at low $p_\mathrm{T}$. Better precision is needed to investigate the expected difference of gluon and light quark energy loss with respect to charm. 
An $R_\mathrm{pPb}(p_\mathrm{T})$ consistent with unity was also measured with ALICE for prompt D mesons with $p_\mathrm{T}$ $<$ 3 GeV/c, confirming that the suppression observed in central \mbox{Pb--Pb} collisions is predominantly induced by final state effects due to charm quark energy loss in the medium \cite{DMesonRpPB}. 
In the right panel of Figure \ref{fig:figure2} the nuclear modification factor of prompt D mesons in the transverse momentum region 8 $<$ $p_\mathrm{T}$ $<$ 16 GeV/{\it c} is shown as a function of centrality. Less suppression is observed moving from central to semi-central collisions, as the medium formed in peripheral collisions should be less dense with respect to the one formed in a central collision.
The comparison of the centrality dependence of the $R_\mathrm{AA}$ of D mesons and of J/$\Psi$ from B-hadron decays (measured by the CMS collaboration \cite{RAACMS}) is displayed in the right panel of Figure \ref{fig:figure2}. It shows an indication for a stronger suppression for charm than for beauty at high $p_\mathrm{T}$ in central Pb--Pb collisions consistent with the expectation of $R^{D}_\mathrm{AA}$ $<$ $R^{B}_\mathrm{AA}$. The $p_\mathrm{T}$ range 8-16 GeV/{\it c} was chosen for D mesons in order to have a similar average transverse momentum (about 10 GeV/{\it c}) than that of B hadrons decaying in a J/$\Psi$ in the measured $p_\mathrm{T}$ interval of 6.5-30 GeV/{\it c}.  The two measurements are described by the predictions based on a pQCD model including mass-dependent radiative and collisional energy loss \cite{Djordjevic}. In this model the difference in $R_\mathrm{AA}$ of charm and beauty mesons is mainly due to the mass dependence of quark energy loss, as demonstrated by the curve in which the non-prompt J/$\Psi$ $R_\mathrm{AA}$ is calculated assuming that b quarks suffer the same energy loss as c quarks.

\begin{figure}[htb!]
\centering
\includegraphics[height=3in]{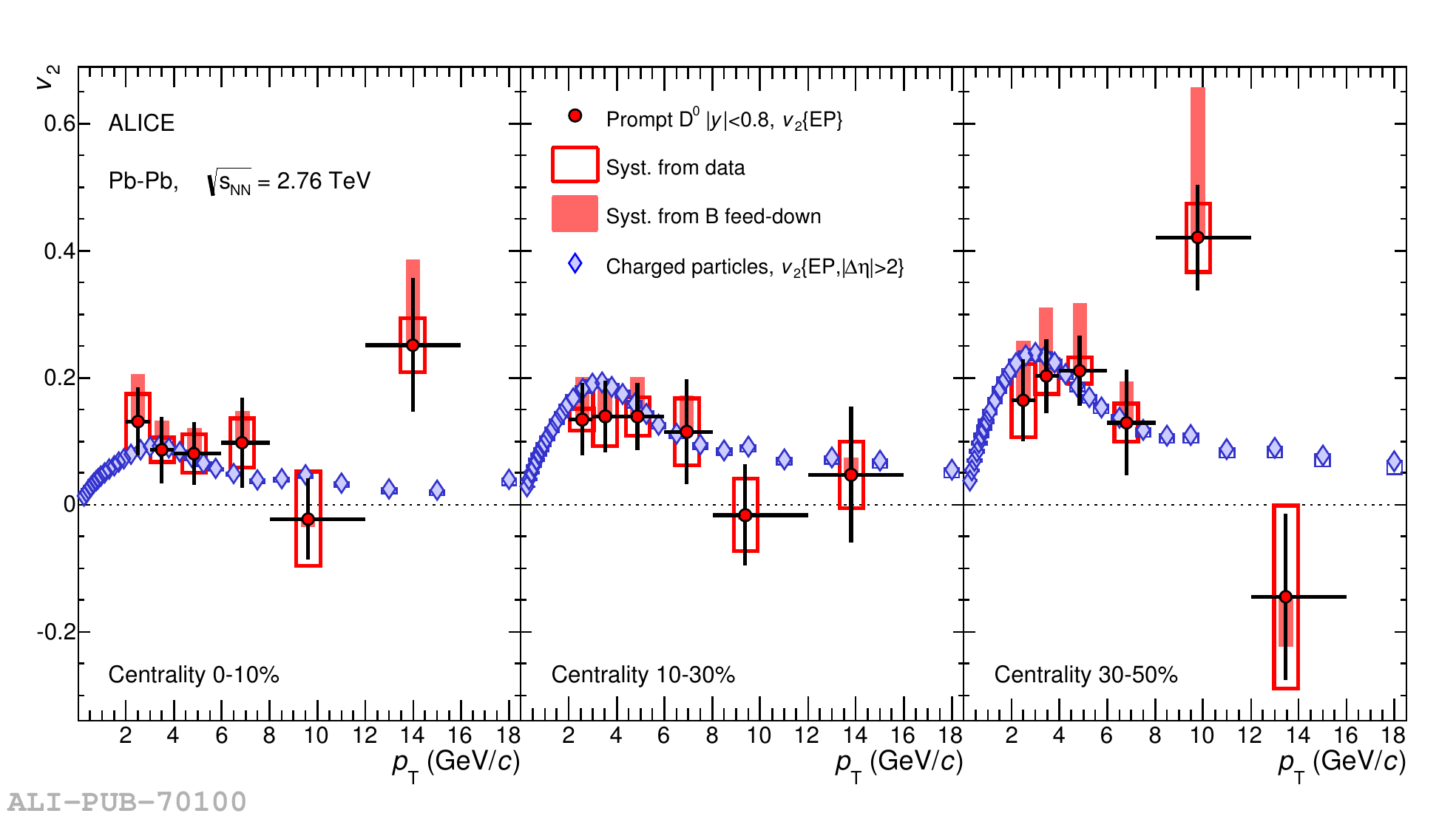}
\caption{ Elliptic flow of $D^{0}$ mesons and charged particles in central and semi-central Pb--Pb collisions at $\sqrt{s_\mathrm{NN}}$  = 2.76 TeV  \cite{DMESONV2}.}
\label{fig:figure7}
\end{figure}

The ALICE collaboration has measured the elliptic flow $v_{2}$ of open heavy-flavour hadrons via their hadronic and semi-leptonic decays in Pb-Pb collisions at $\sqrt{s_\mathrm{NN}}$ = 2.76 TeV.
The measured averaged $v_{2}$  of prompt $D^{0}$, $D^{+}$ and $D^{*+}$  indicates a positive $v_{2}$ in semi-central (30-50\%) Pb--Pb collisions with a significance of 5.7$\sigma$ for 2 $<$ $p_\mathrm{T}$ $<$ 6 GeV/{\it c} \cite{DMESONV2}.
The anisotropy of prompt  $D^{0}$ mesons was measured in the three centrality classes 0-10\%, 10-30\% and 30-50\%, as reported in Figure \ref{fig:figure7}. The results show a hint of increasing $v_{2}$ from central to semi-peripheral collisions and are comparable in magnitude to that of inclusive charged particles, suggesting that charm quarks participate in the collective flow of the expanding medium.

 \begin{figure}[htb!]
\begin{minipage}[b]{0.334\linewidth}
\centering
\includegraphics[height=2.34in]{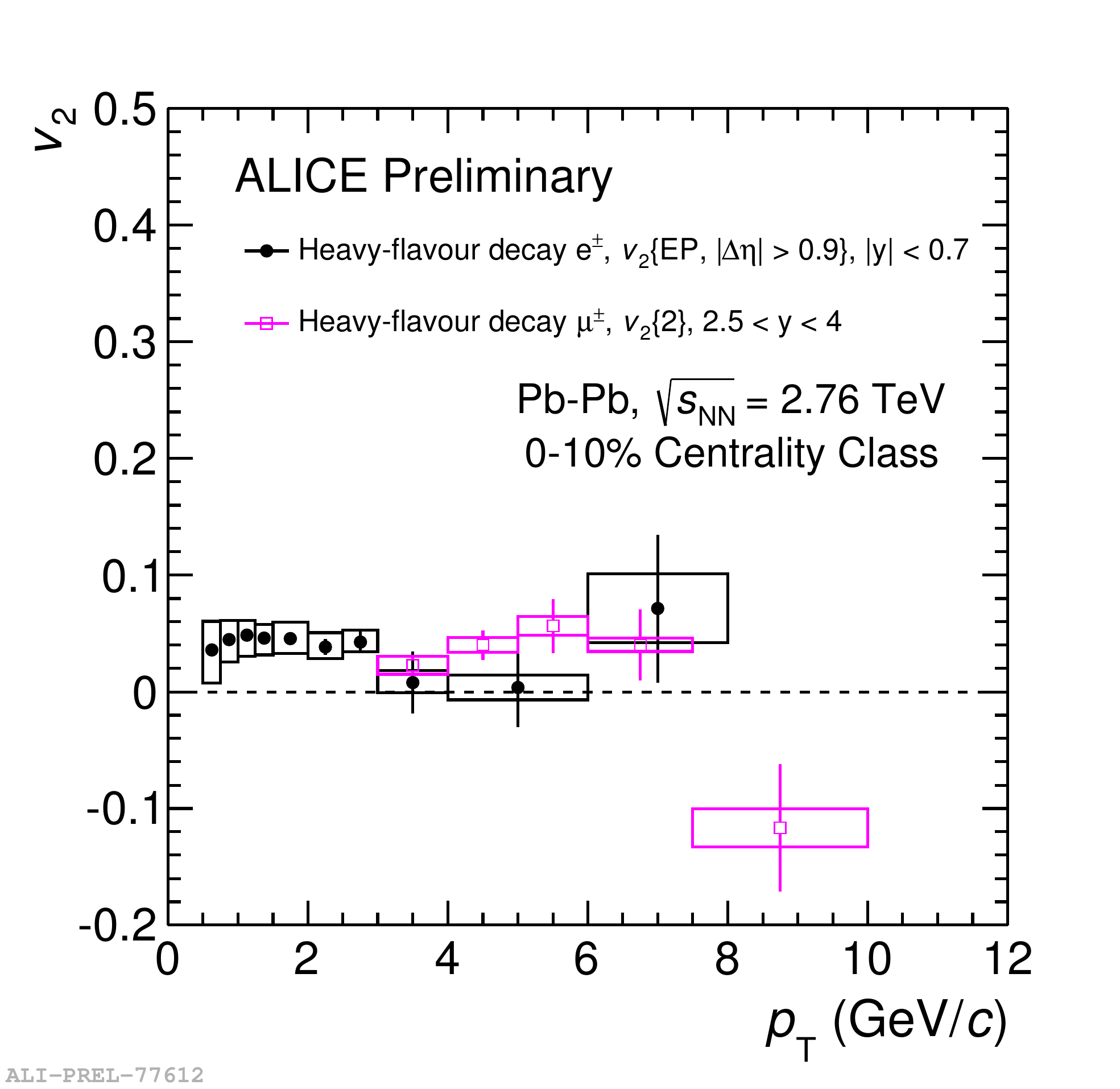}
\end{minipage}
 \begin{minipage}[b]{0.334\linewidth}
\centering
\includegraphics[height=2.34in]{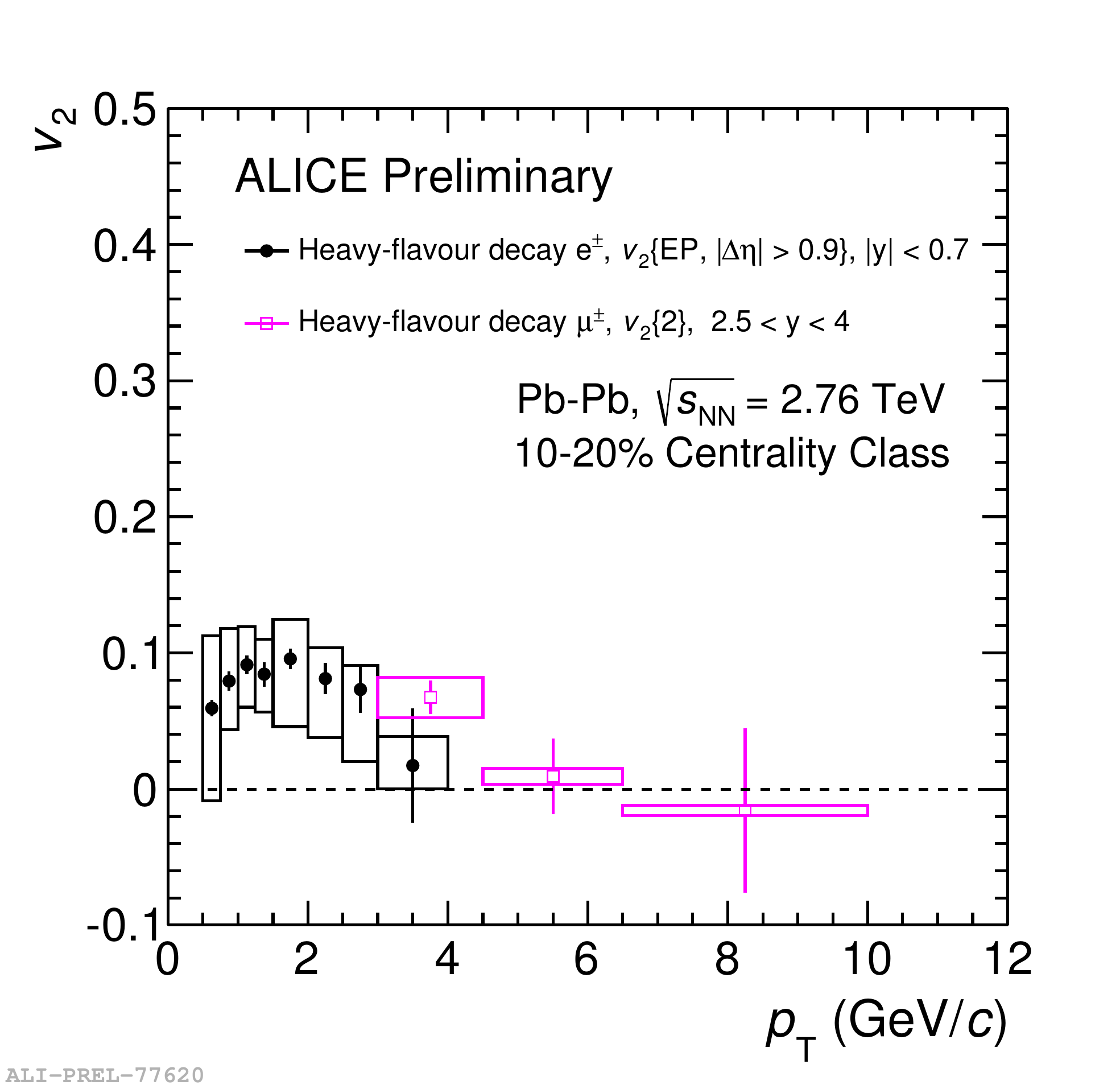}
\end{minipage}
 \begin{minipage}[b]{0.0\linewidth}
\centering
\includegraphics[height=2.34in]{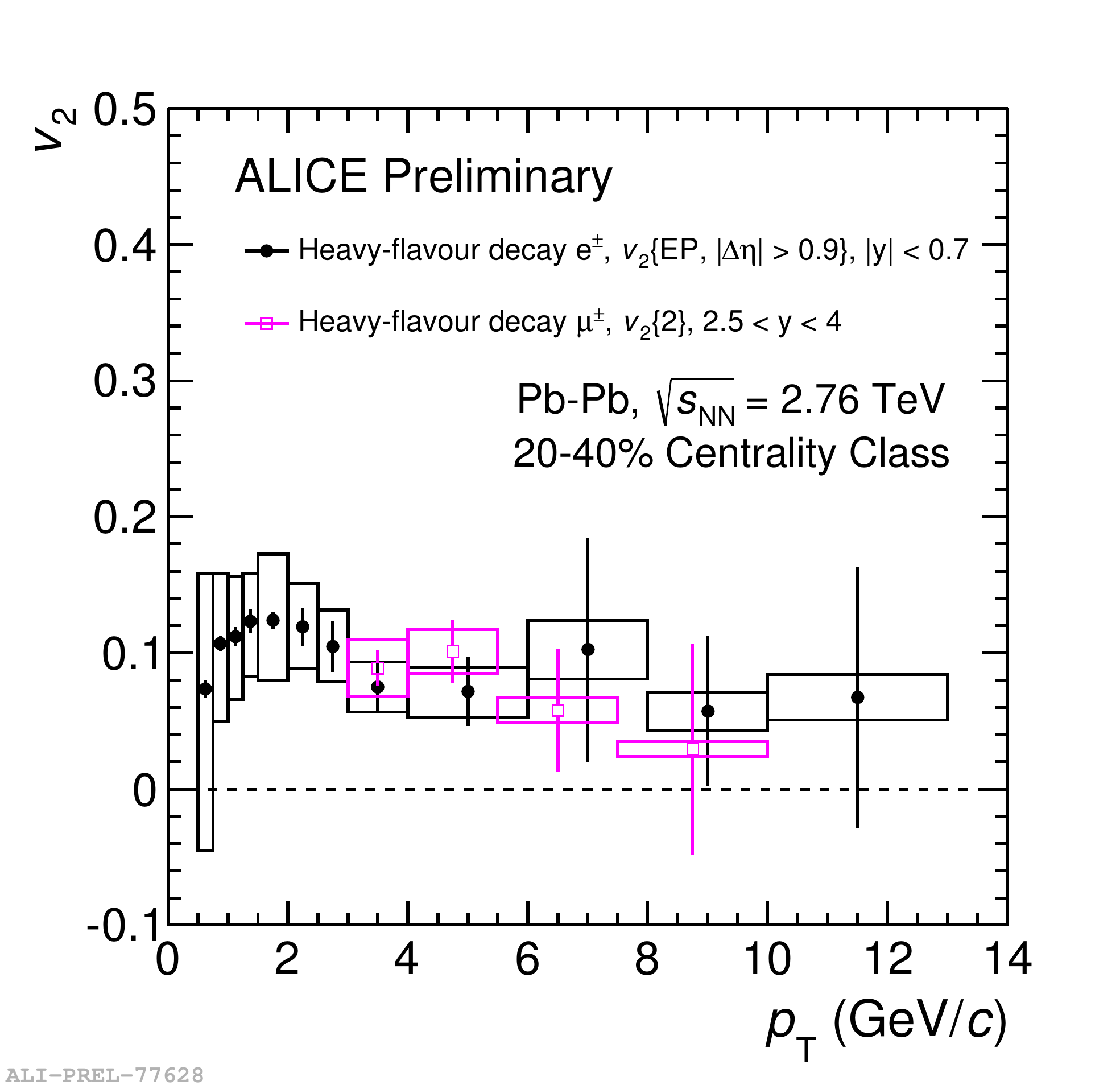}
\end{minipage}
\caption{Elliptic flow of heavy-flavour decay electrons at mid-rapidity and heavy-flavour decay muons at forward rapidity measured with the event-plane method and with two-particle Q-cumulant, respectively, in Pb--Pb collisions at $\sqrt{s_\mathrm{NN}}$ = 2.76 TeV for three centrality intervals.}
\label{fig:figure8}
\end{figure}

In order to determine the elliptic flow of heavy-flavour decay electrons, the contribution from background electrons needs to be subtracted from the elliptic flow of inclusive electrons. 
The electron background, mainly originates from photonic sources, namely conversions of photons in the detector material and Dalitz decays of $\pi^{0}$ and $\eta$ mesons. The raw yield and $v_{2}$ of photonic electrons was measured by pairing electrons with opposite-charge partner tracks and requiring a small invariant mass of the reconstructed $\mathrm{e}^{+}\mathrm{e}^{-}$ pair. At low $p_\mathrm{T}$, the inclusive electron sample is dominated by background electrons, whereas at high $p_\mathrm{T}$ electrons from heavy-flavour decays give the main contribution. For $p_\mathrm{T}$ $>$1.5 GeV/{\it c}, the $v_{2}$ measurement of the background sources starts to be limited by statistics. Therefore a cocktail approach is used, where the measured $v_{2}$ and $p_\mathrm{T}$ spectra of $\pi^{0}$, $\pi^{\pm}$ and direct $\gamma$ were employed as input.
The elliptic flow of heavy-flavour decay electrons was measured in the three centrality classes 0-10\%, 10-20\% and 20-40\% (see Figure \ref{fig:figure8}). The magnitude of $v_{2}$ increases from central to semi-central collisions. In semi-central (20-40\%) Pb--Pb collisions a positive $v_{2}$ is observed with a significance of 3$\sigma$ for 2 $<$ $p_\mathrm{T}$ $<$ 3 GeV/{\it c}. This confirms the significant interaction of heavy quarks with the medium. 
The $v_{2}$ of heavy-flavour decay muons is measured in the three centrality classes 0-10\%, 10-20\% and 20-40\% with the two-particle Q-cumulant method (see Figure \ref{fig:figure8}).
The analysis strategy is similar to the one for heavy-flavour decay electrons. The elliptic flow of heavy-flavour decay muons is obtained by subtracting the background contribution, mainly due to $\pi^{\pm}$ and $K^{\pm}$ decays, from the measured elliptic flow of inclusive muons. The $v_{2}$ of $\pi^{\pm}$ and $K^{\pm}$ decay muons is estimated by extrapolating to forward rapidity the elliptic flow of charged hadrons measured in $|\eta|$ $<$ 2.5 by the ATLAS collaboration \cite{ATLAS}. 
A positive $v_{2}$ is observed in semi-central (20-40\%) Pb--Pb collisions with a significance larger than 3$\sigma$ for 3$<$ $p_\mathrm{T}$ $<$ 5 GeV/{\it c}. The results are comparable in magnitude with the measured elliptic flow of heavy-flavour decay electrons at mid-rapidity.


\section{Model comparison}

%

\begin{figure}[htb!]
\begin{minipage}[b]{0.5\linewidth}
\centering
\includegraphics[height=2.8in]{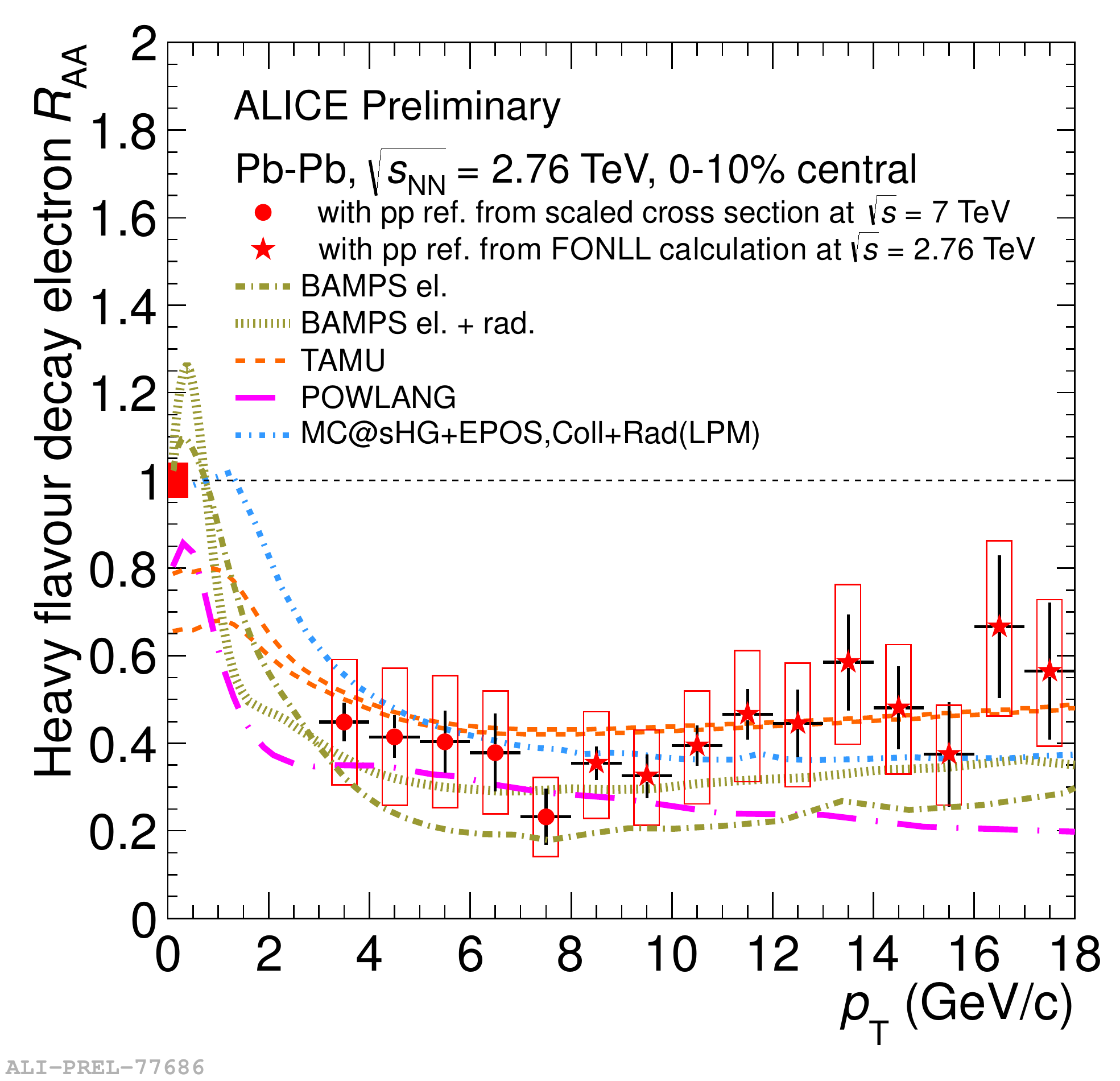}
\end{minipage}
\hspace{0.1cm}
\begin{minipage}[b]{0.5\linewidth}
\centering
\includegraphics[height=3.in]{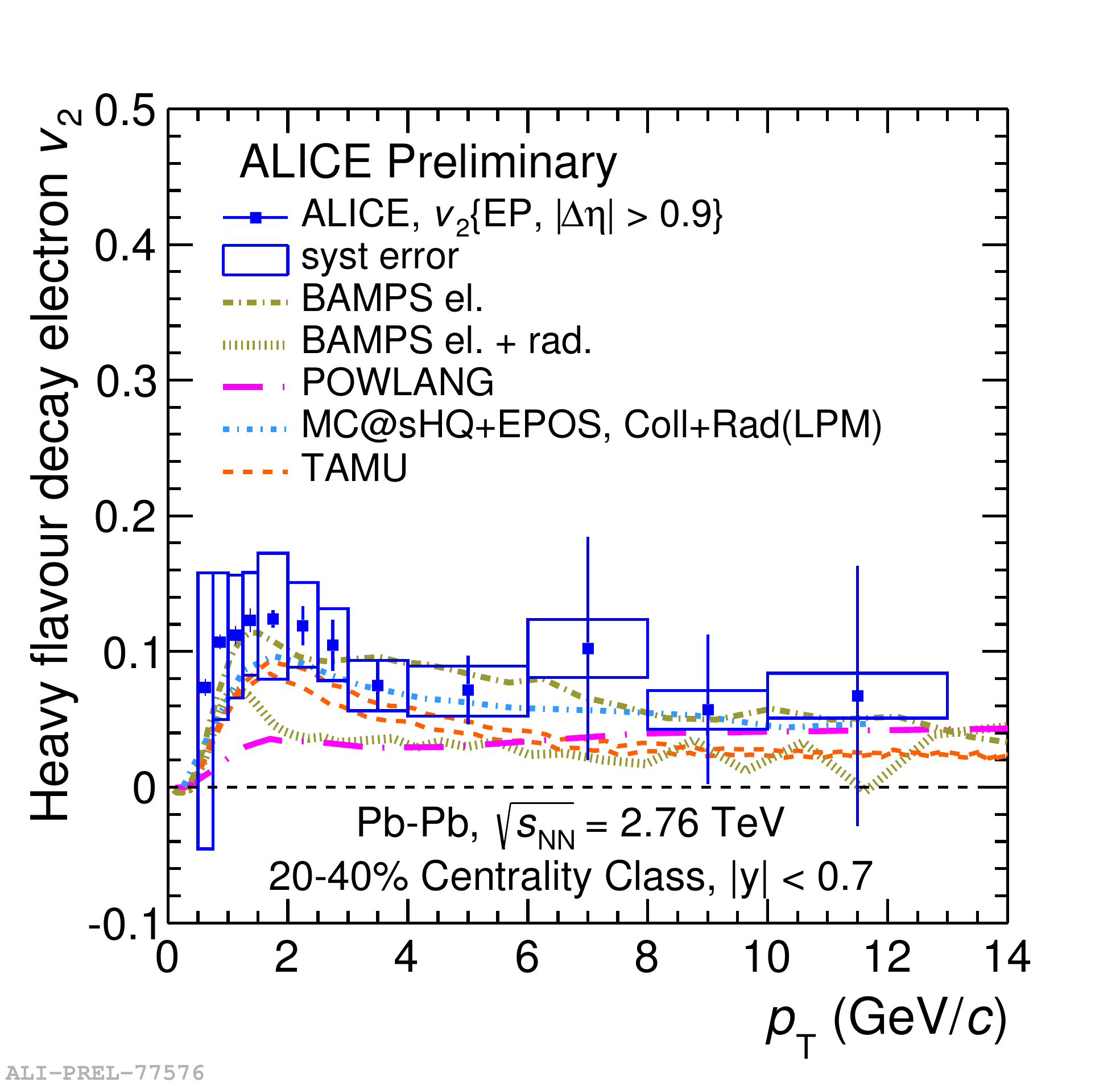}
\end{minipage}
\caption{ 
Comparison of heavy-flavour decay electron $R_\mathrm{AA}$ in central Pb--Pb collisions (left panel) and $v_{2}$ in semi-peripheral collisions (right panel) to model calculations \cite{BAMPS1}, \cite{BAMPS2}, \cite{TAMU},  \cite{POWLANG},  \cite{EPOS}.}
\label{fig:figure10}
\end{figure}

Several theoretical model calculations are available for the elliptic flow coefficient $v_{2}$ and the nuclear modification factor $R_\mathrm{AA}$ of heavy-flavour hadrons and their decay leptons.
In Figure \ref{fig:figure10} the measured heavy-flavour decay electron $R_\mathrm{AA}$ for the 0-10\% centrality class (left) and $v_{2}$ for the 20-40\% centrality class  (right) are compared to different parton transport models: BAMPS, which includes collisional in-medium energy loss and mimics the radiative processes by increasing the elastic cross section  \cite{BAMPS1}; BAMPS el. + rad., which includes radiative processes  \cite{BAMPS2}; TAMU, which incorporates energy loss via collisional processes, with resonance formation and dissociation in an evolving hydrodynamic medium  \cite{TAMU}; POWLANG, which is based on the Langevin transport equation with collisional energy loss in an expanding, deconfined medium \cite{POWLANG}; and MC@sHG+EPOS, which includes radiative and collisional energy loss in an expanding medium based on the EPOS model  \cite{EPOS}. 
The models including interactions of the c and b quarks with a hot, dense and deconfined medium can qualitatively describe the features observed in the data.
A simultaneous description of the  $R_\mathrm{AA}$ and $v_{2}$ starts to provide constraints to the models themselves.



\section{Conclusions}

The results obtained with ALICE using the data from the LHC Run-1 (2010-2013) indicate a strong suppression of heavy-flavour production in central Pb--Pb collisions for $p_\mathrm{T}$ $>$ 3 GeV/c, observed for heavy-flavour decay electrons and muons, for electrons from beauty-hadron decays and for prompt D mesons. From the comparison with p--Pb measurements it is possible to conclude that the suppression observed in Pb--Pb collisions is mainly due to final state effects, i.e. the interaction of heavy quarks with the hot and dense medium.
Finally the centrality dependent D-meson $R_\mathrm{AA}$, measured in the transverse momentum range \mbox{8 $<$ $p_\mathrm{T}$ $<$16 GeV/{\it c}}, was compared with the $R_\mathrm{AA}$ of non-prompt J/$\Psi$ measured by the CMS collaboration and a clear indication of a different suppression was found as expected from the anticipated quark-mass dependence of the energy loss.
The azimuthal anisotropy of D mesons and heavy-flavour decay electrons at mid-rapidity, as well as heavy-flavour decay muons at forward rapidity, measured with ALICE in central and semi-central Pb--Pb collisions at $\sqrt{s_\mathrm{NN}}$ = 2.76TeV was presented. The results indicate a positive $v_{2}$  with a significance larger than 3$\sigma$ in semi-central Pb--Pb collisions and a hint that $v_{2}$ increases  from central to semi-central collisions. The magnitude of the D-meson elliptic flow is similar to the one of charged particles. This suggests a collective motion of low-$p_\mathrm{T}$ heavy-quarks, mainly charm. The $v_{2}$  and $R_\mathrm{AA}$ measurements together can provide constraints to the models.


\end{document}